\documentclass[iop,numberedappendix,apj]{emulateapj}

\usepackage{epsfig}
\usepackage{amsmath,amsthm,amssymb,cases}
\usepackage{rotating}
\usepackage{natbib}
\usepackage{enumerate}
\usepackage{multirow}
\usepackage{array}
\usepackage{appendix}
\usepackage{comment}
\usepackage{color,xcolor}
\usepackage{url}
\usepackage{ulem}
\usepackage{here}
\usepackage{hyperref}
\usepackage{bm}
\hypersetup{colorlinks,linkcolor={blue!50!black},citecolor={blue!50!black},urlcolor={blue!50!black}}
\allowdisplaybreaks[1]
\renewcommand{\bibname}{References}

\def\gsim{\;\rlap{\lower 2.5pt
 \hbox{$\sim$}}\raise 1.5pt\hbox{$>$}\;}
\def\lsim{\;\rlap{\lower 2.5pt
   \hbox{$\sim$}}\raise 1.5pt\hbox{$<$}\;}
\def\ftilde{\tilde f}

\begin{document}

\title{Rotational Spectral Unmixing of Exoplanets:\\Degeneracies between Surface Colors and Geography}

\author{
Yuka Fujii\altaffilmark{1,2} 
Jacob Lustig-Yaeger\altaffilmark{3,4,5} 
Nicolas B. Cowan\altaffilmark{6,7} 
}

\affil{$^1$NASA Goddard Institute for Space Studies, 
  New York, NY 10025, USA}
      
\affil{$^2$Earth-Life Science Institute, Tokyo Institute of Technology, 
  Tokyo, 152-8550, JAPAN}
  
\affil{$^3$Astronomy Department, University of Washington, Box 951580, Seattle, WA 98195, USA}

\affil{$^4$Astrobiology Program, University of Washington, 3910 15th Ave. NE, Box 351580, Seattle, WA 98195, USA}

\affil{$^5$NASA Astrobiology Institute -- Virtual Planetary Laboratory Lead Team, USA}

\affil{$^6$Department of Earth and Planetary Sciences, McGill University, Montreal, Quebec, H3A 0E8, Canada}

\affil{$^7$Department of Physics, McGill University, Montreal, Quebec, H3A 0E8, Canada}

\vspace{0.5\baselineskip}

\email{
yuka.fujii.ebihara@gmail.com
}

\begin{abstract}
Unmixing the disk-integrated spectra of exoplanets provides hints about heterogeneous surfaces that we cannot directly resolve in the foreseeable future. 
It is particularly important for terrestrial planets with diverse surface compositions like Earth. 
Although previous work on unmixing the spectra of Earth from disk-integrated multi-band light curves appeared successful, we point out a mathematical degeneracy between the surface colors and their spatial  distributions. 
Nevertheless, useful constraints on the spectral shape of individual surface types may be obtained from the premise that albedo is everywhere between 0 and 1.
We demonstrate the degeneracy and the possible constraints using both mock data based on a toy model of Earth, as well as real observations of Earth. 
Despite the severe degeneracy, we are still able to recover an approximate albedo spectrum for an ocean. 
In general, we find that surfaces are easier to identify when they cover a large fraction of the planet and when their spectra approach zero or unity in certain bands. 
\end{abstract}

  

\section{Introduction}
\label{sec:intro}

Direct imaging of exoplanets is expected to play a vital role in characterizing Earth analogs in habitable zones and beyond. 
Substantial work has gone into predicting detectable features in disk-integrated spectra of Earth and other planets, as they are observed from an astronomical distance. 
Disk-integrated spectra or even multi-band photometry (colors) in principle include the effects of the surface reflectance as well as the atmosphere. (We use the terms spectra and colors interchangeably.) 
However, interpreting disk-integrated colors is not trivial. 
This is particularly true for Earth-like planets that harbor diverse atmospheric and surface characteristics, including liquid water, partial cloud cover, continents, and possibly vegetation. 

A key here is to leverage the time variation of the spectrum/color  \citep{Ford2001}: the regions that contribute to the scattered light change due to the planetary rotation and  orbital motion, so the time variability can be used to map the heterogeneity of the surface environment \citep[for a recent review, see][]{Cowan2017}.  
\citet{Cowan2009, Cowan2011} performed principal component analysis (PCA) on the observed multi-band photometry of Earth. They found that the number of surface types can be inferred from the number of dominant principal components: (\# of surface types) $\ge $ (\# of principal components) $+ 1$.  
\citet{Fujii2010, Fujii2011} decomposed multi-band photometric data of Earth assuming the template reflectance spectra of the known major surface types, and they found that the relative abundance and longitudinal variation of these surfaces are approximately recovered. 
Moreover, by coupling the rotational variation with orbital (phase) variation, a two-dimensional map of the surface may be retrieved \citep{Kawahara2010, Kawahara2011, Fujii2012}. 

\citet{Cowan2013} took another approach to the same inverse problem. 
Their strategy was to estimate the reflectance spectra of surfaces and their geographical maps across the globe simultaneously, by making all of them fitting parameters. 
They applied their method to the light curves of Earth observed by EPOXI, and appeared successful in that they obtained reflectance spectra roughly matching the average spectra of clouds, oceans, and continents. 
However, the longitudinal map of these components did not match the actual geography.  

Motivated by this unsatisfactory result, we revisited their analysis. 
We start by creating mock light curves to see if we can fully retrieve the known answer. 
Taking a close look at the mathematical relation between the colors and geographical maps of the surfaces, we discuss the degeneracy among the parameters to be estimated. 
We then demonstrate, using mock data, how our estimates are nevertheless constrained by boundary conditions, including the expectation that albedo is between 0 and 1. 
We also apply the same procedure to the EPOXI data. 

This paper is organized as follows. 
In Section \ref{s:mockdata}, we introduce our toy-model multi-band diurnal light curves of Earth. 
We analyze the mock data in Section~\ref{s:frame} and develop a general framework for this problem. 
In Section \ref{s:EPOXI}, we apply our procedure to the EPOXI light curves and attempt to estimate the surface colors of Earth. 
In Section \ref{s:discussion}, we discuss possible improvements and a few confounding factors, including the degeneracy between the planetary albedo and radius. 
We conclude in Section~\ref{s:conclusion}. 

\begin{figure}[h]
    \begin{center}
	\includegraphics[width=\hsize]{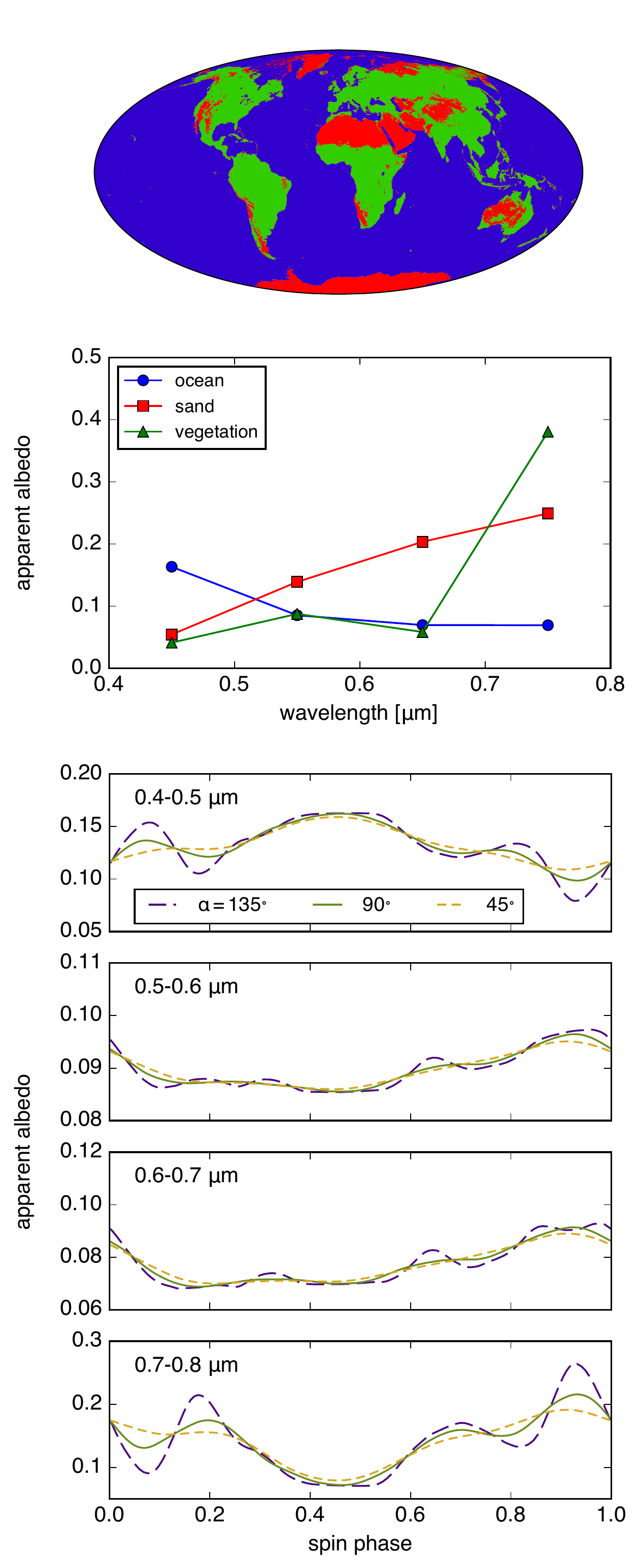}
    \end{center}
    \caption{Our mock data based on the IGBP classification map of Earth. \emph{Top panel:} distribution of three surface types (ocean: blue; sand: red; vegetation: green). \emph{Next panel:} assumed albedo spectra with matching colors. \emph{Bottom four panels:} rotational light curves in four  photometric band with varying phase angles, $\alpha = 135^{\circ }$ (purple, long-dashed), $90^{\circ }$ (olive, solid), and $45^{\circ }$ (gold, short-dashed). }
\label{fig:mockdata}
\end{figure}

\section{Preparing Mock Data Sets}
\label{s:mockdata}

In order to facilitate the discussions in the following sections, in this section we introduce a mock dataset to be used for demonstrations. 

We consider diurnal light curves of a toy model of the atmosphere-less Earth, in four photometric bands. 
We use a simplified surface map as shown in the top panel of Figure \ref{fig:mockdata}. 
This map is based on the land classification scheme of the International Geosphere-Biosphere Programme (IGBP)\footnote{\url{https://climatedataguide.ucar.edu/climate-data/ceres-igbp-land-classification}}. 
Although the original classification assumes 16 land surface types plus ocean, in this paper we adopt three surface types for simplicity regarding ``Open Shrublands,'' ``Permanent Wetlands,'' ``Urban,'' ``Snow/Ice,'' and ``Barren/Desert'' as ``sand'' (red in the top panel of Figure \ref{fig:mockdata}) and other land surface types as ``vegetation'' (green), while keeping the ``ocean'' (blue) regions. 

The assumed albedo spectra of these surface types in four photometric bands are shown in the middle panel of Figure \ref{fig:mockdata}. 
These four photometric bands correspond to 0.4--0.5 $\mu $m, 0.5--0.6 $\mu $m, 0.6--0.7 $\mu $m, and 0.7--0.8 $\mu $m, respectively, but we will simply refer to them by the band indices ($j$) unless otherwise noted. 
The albedo spectrum for an ocean is based on \citet{Mclinden1997}, 
and the data for sand and vegetation are taken from the ASTER spectral library\footnote{\url{https://speclib.jpl.nasa.gov/}. Specifically, we adopt  ``Brown to dark brown sand'' for ``sand,'' and ``Grass'' for the ``vegetation.''}. 
Scattering from the surface is assumed to obey the Lambert law, i.e., the radiance is (incident flux)$\times $(albedo)/$\pi$, independent of the direction of  scattering. 
Note that in reality they are not Lambertian scatterers; among others,  scattering by the ocean is particularly anisotropic at the crescent phase (see Section \ref{ss:deviate_Lambert}).  

The diurnal light curves are synthesized given the relative positions of the star, planet, and observer. 
For the sake of simplicity, we consider a planet with zero obliquity in an edge-on orbit, and change the phase angle (the planet-centric angle between the star and the observer) denoted by $\alpha $ to $45^{\circ }$, $90^{\circ }$, and $135^{\circ }$. 
We also assume that the spin period is significantly shorter than the orbital period and that the orbital phase does not change in a single spin rotation. 
However, the following discussion does not depend on these assumptions. 
We consider noise-free data. Clearly, noise in real observations of Earth twins is expected to be substantial, and the effect of such observations will be discussed in a forthcoming study. 

The bottom panels of Figure \ref{fig:mockdata} display examples of diurnal light curves in four photometric bands, represented in terms of {\it apparent albedo}. 
Apparent albedo is the albedo of the planet weighted by illumination and visibility; it is obtained by normalizing the planetary intensity by that of a lossless Lambert sphere with the same radius and at the same phase \citep{Qiu2003, Seager2010}. 
In this paper, we use apparent albedo unless otherwise noted. 
The apparent albedo is straightforwardly obtained from the observed planetary intensity, if and only if the orbital phase of the planet, the distance between the star and the planet, and the planetary radius are all known. 
We assume these quantities are known, but discuss the effect of unknown planetary radius in Section \ref{sss:uncertain_radius}.

The problem throughout this paper is that, from these kinds of multi-band diurnal light curves, we would like to know how---and how well---we can retrieve the albedo spectra of different surface types and the longitudinal distributions of these surface types.

\section{Inverse problem}
\label{s:frame}

In this section, we discuss the general framework for analyzing the diurnal light curves, and present some demonstrations using mock data created in the previous section. 
Sections \ref{ss:model} and \ref{ss:PCplane} are essentially the recapitulation of previous papers, in particular \citet{Cowan2013} \citep[see also][]{Cowan2009,Cowan2011,Fujii2010,Fujii2011}.  
We choose to include these discussions, however, as a baseline to establish the later arguments. 

\subsection{Algebraic Formulation}
\label{ss:model}

\begin{table}[b]
\caption{Indexes}
\begin{center}
\begin{tabular}{lcc} \hline \hline
Name & Symbol & Maximum \\ \hline
Observation Time & $i$ & $I$ \\
Band & $j$ & J  \\
Surface Type & $k$ & $K$  \\
Longitudinal Slice  & $l$ & $L$ \\ 
Principal Components & $n$ & $N$ \\ \hline
\end{tabular}
\end{center}
\label{tab:index}
\end{table}%

Assuming that the planetary surface is everywhere Lambertian scattering, and that it is composed of a certain number $K$ of spectrally distinct surface types, then the disk-integrated scattered light is a weighted summation of the reflectance spectra of different surface types, as shown below. 

The apparent albedo of the planet at time $t$ and at wavelength $\lambda $, $d (t, \lambda )$ (``$d$'' for data), is the integral of the local albedo (denoted by $s (\lambda, \bm{\Omega })$, where $\bm{ \Omega }$ indicates a position on the planetary surface), weighted by the cosine of the zenith angle of insolation (denoted by $\theta _0 (t, {\bm \Omega})$) and the cosine of the zenith angle of the observer (denoted by $\theta _1 (t, {\bm \Omega})$), i.e.,:
\begin{eqnarray}
d (t_i, \lambda_j) &=& \displaystyle \frac{ \int_{{\rm IV}_i} s (\lambda _j, {\bm \Omega }) \, W (t_i, {\bm \Omega}) \; d \Omega }{ \int_{{\rm IV}_i}  W (t_i, {\bm \Omega}) \; d \Omega }  \\
W (t, {\bm \Omega}) &\equiv & \cos \theta_0 (t, {\bm \Omega}) \cos \theta_1 (t, {\bm \Omega}) 
\end{eqnarray}
where $i$ and $j$ are the indices for time and wavelength, respectively, and ${\rm IV }_i$ represents the illuminated and visible area of the planetary surface.
The weight function, $W (t, {\bm \Omega})$, is also referred to as the convolution kernel \citep{Cowan2017}.
In the following, the maximum values of $i$ and $j$ are denoted by $I$ and $J$, respectively, as summarized in Table \ref{tab:index}, along with other related indices.

By decomposing the local albedo as $s(\lambda _j, {\bm \Omega }) = \sum _k s_{kj} f_k({\bm \Omega })$ where $s_{kj}$ and $f_k({\bm \Omega })$ are the spectrum and the local area fraction of $k$-th surface type, respectively, the above equation becomes
\begin{eqnarray}
d (t_i, \lambda_j) &=& \sum _{k} s_{kj} \; \displaystyle \frac{ \int_{{\rm IV}_{i}} f_k (\bm \Omega ) \, W (t_i, {\bm \Omega}) \; d \Omega }{ \int_{{\rm IV}_i}  W (t_i, {\bm \Omega}) \; d \Omega } \notag \\
&=& \sum _{i,k} \ftilde_{ik} \, s_{kj} \label{eq:tilde_d_f_ast_s}
\end{eqnarray}
where $\tilde f_{ik}$ represents the apparent covering fraction of $k$-th surface type at time $t_i$:
\begin{equation}
\tilde f_{ik} \equiv  \frac{ \int_{{\rm IV}_{i}} f_k (\bm \Omega ) W (t_i, {\bm \Omega}) \; d \Omega }{ \int_{{\rm IV}_i}  W (t_i, {\bm \Omega}) \; d \Omega } \label{eq:def_ftilde}
\end{equation}
Thus, spectral unmixing is equivalent to estimating both the apparent covering fraction ($\ftilde_{ik}$) and colors ($s_{kj}$) from the apparent albedo of the planet ($d (t_i, \lambda_j)\equiv d_{ij}$).

Because $f_k({\bm \Omega })$ should not be negative and should sum up to unity, $\ftilde _{ik}$ should also not be negative and they should sum up to unity at each observation time, $i$. In addition, reflectance spectra, $s_{kj}$, should be between 0 and 1 for any surface type ($k$) at any band ($j$). 
Therefore, the constraints on apparent covering fraction ($\ftilde_{ik}$) and colors ($s_{kj}$) of the surfaces are:
\begin{subnumcases}
{}
0 \leq \ftilde_{ik} \;\;\; & \mbox{for any $i$, $k$} \label{eq:tilde_f_range} \\
\sum_k \ftilde_{ik} = 1 & \mbox{for any $i$} \label{eq:tilde_f_sum} \\
0 \leq s_{kj} \leq 1 \;\;\; & \mbox{for any $k$, $j$} \label{eq:tilde_s_range} 
\end{subnumcases}

The time variability of the apparent covering fraction, $\ftilde $, due to the planet's rotation, is related to the surface inhomogeneity in the longitudinal direction. 
From Equation (\ref{eq:def_ftilde}), $\ftilde _{ik}$ may be approximately written as the weighted sum of the average area fraction of the $k$-th surface type in the $l$-th longitudinal slice, denoted by $f_{lk}$:
\begin{eqnarray}
\ftilde _{ik} &=& \frac{ \int_{{\rm IV}_i} f_k (\bm \Omega ) \; W (t_i, {\bm \Omega})  \; d\Omega }{ \int_{{\rm IV}_i}  W (t_i, {\bm \Omega}) \; d\Omega }  \\
&\approx & \sum_l  W_{il} f_{lk} \label{eq:Wf}, \\
W_{il} &\equiv & \frac{ \int_{{\rm IV}_i \cap \Omega _l}  W (t_i, {\bm \Omega}) \; d\Omega }{ \int W (t_i, {\bm \Omega})  \; d\Omega }
\end{eqnarray}
where $\Omega _l$ denotes the $l$-th longitudinal slice.
Strictly speaking, the approximation is valid only when the local area fraction, $f_k(\bm{\Omega })$, does not change or changes little across each longitudinal slice for all $K$ surface types. 

One can therefore recast Equation (\ref{eq:tilde_d_f_ast_s}) in terms of the longitudinal map, $f_{lk}$:
\begin{equation}
d_{ij} \approx \sum _{l,k} W_{il} \, f_{lk} \, s_{kj} \label{eq:d_f_s}
\end{equation}
Again, the area fractions in each longitudinal slice, $f_{lk}$, cannot be negative and should sum up to unity. Thus, a set of conditions similar to Equations (\ref{eq:tilde_s_range})-(\ref{eq:tilde_f_sum}) is imposed:
\begin{subnumcases}
{}
0 \leq f_{lk} \;\;\; & \mbox{for any $l$, $k$} \label{eq:f_range} \\
\sum_k f_{lk} = 1 & \mbox{for any $l$} \label{eq:f_sum} \\
0 \leq s_{kj} \leq 1 \;\;\; & \mbox{for any $k$, $j$} \label{eq:s_range}
\end{subnumcases}
Now the relevant problem is to estimate both the geographical map ($f_{lk}$) and colors ($s_{kj}$) given the time series of apparent albedo of the planet ($d_{ij}$), subject to the constraints of Equations (\ref{eq:s_range})-(\ref{eq:f_sum}); this is where \citet{Cowan2013} stood. 
In principle, constraints (\ref{eq:f_range}) and (\ref{eq:f_sum}) for the longitudinal geographical map are more stringent than constraints (\ref{eq:tilde_f_range}) and (\ref{eq:tilde_f_sum}) for the apparent covering fraction, due to the low-pass filter nature of the convolution from map to light curve.

\begin{figure}[b!]
    \begin{center}
\includegraphics[width=0.9\hsize]{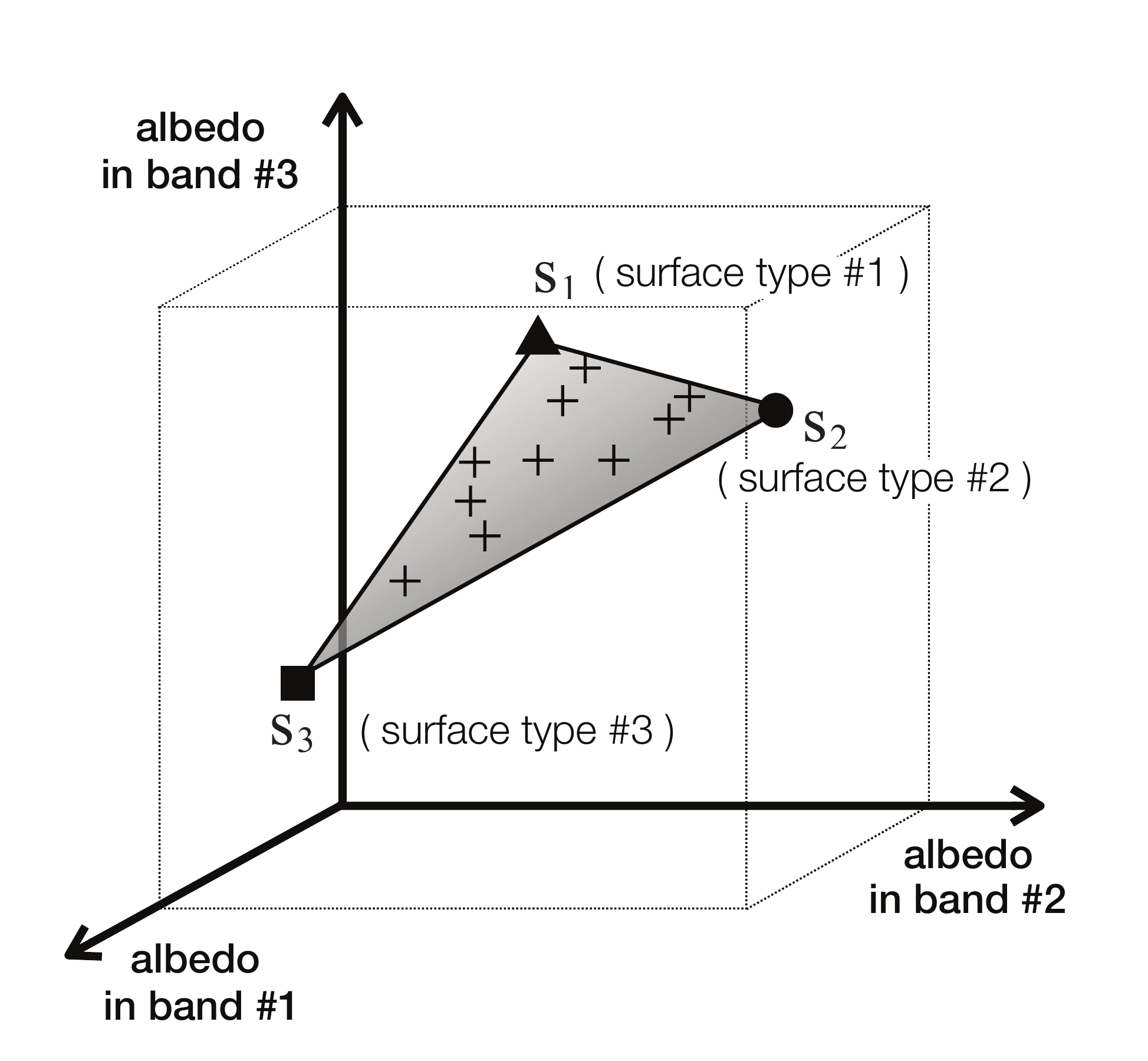}
    \end{center}
    \caption{Schematic figure to illustrate the relation between observed data ($\{{\bm d}_i\} $; crosses) and surface spectra ($\{{\bm s}_k\} $; filled points). The dashed cube shows the range of physical albedos (between 0 and 1). }
\label{fig:schematic}
\end{figure}

\subsection{Graphical Conception of \\Principal Component Plane}
\label{ss:PCplane}

Equation (\ref{eq:tilde_d_f_ast_s}), coupled with the conditions (\ref{eq:tilde_f_range}) and (\ref{eq:tilde_f_sum}), indicates that the observed planetary albedo ($d_{ij}$) is a convex combination of the colors of representative surface types ($s_{kj}$). 
Geometrically, this means that in the $J$-dimensional color space, $\{{\bm d}_i\}$ are located in the convex hull with $K$ vertices, $\{{\bm s}_k\} $. 
For example, in the case of $K=3$, this convex hull is simply a triangle, and thus $\{{\bm d}_i\}$ are confined in the triangular region of a plane. 
Figure \ref{fig:schematic} graphically shows this relation in the case of $K=3$ and $J=3$. 
In general, the data points, $\{{\bm d}_i\}$, are in a $K-1$ dimensional subspace of the $J$-dimensional color space. 

This subspace can be identified by performing PCA \citep{Cowan2009,Cowan2011}, as PCA extracts the major, mutually orthogonal axes along which the data are scattered. 
Consequently, the number of dominant principal components (PCs) is less than or equal to $K-1$ \citep{Cowan2011}. 
Note that non-Lambertian reflectance, partially transparent cloud cover, or observational noise would in practice ensure that the PC space has some additional dimensions (see also Section \ref{ss:deviate_Lambert}). 

\begin{figure}[t]
    \begin{center}
\includegraphics[width=\hsize]{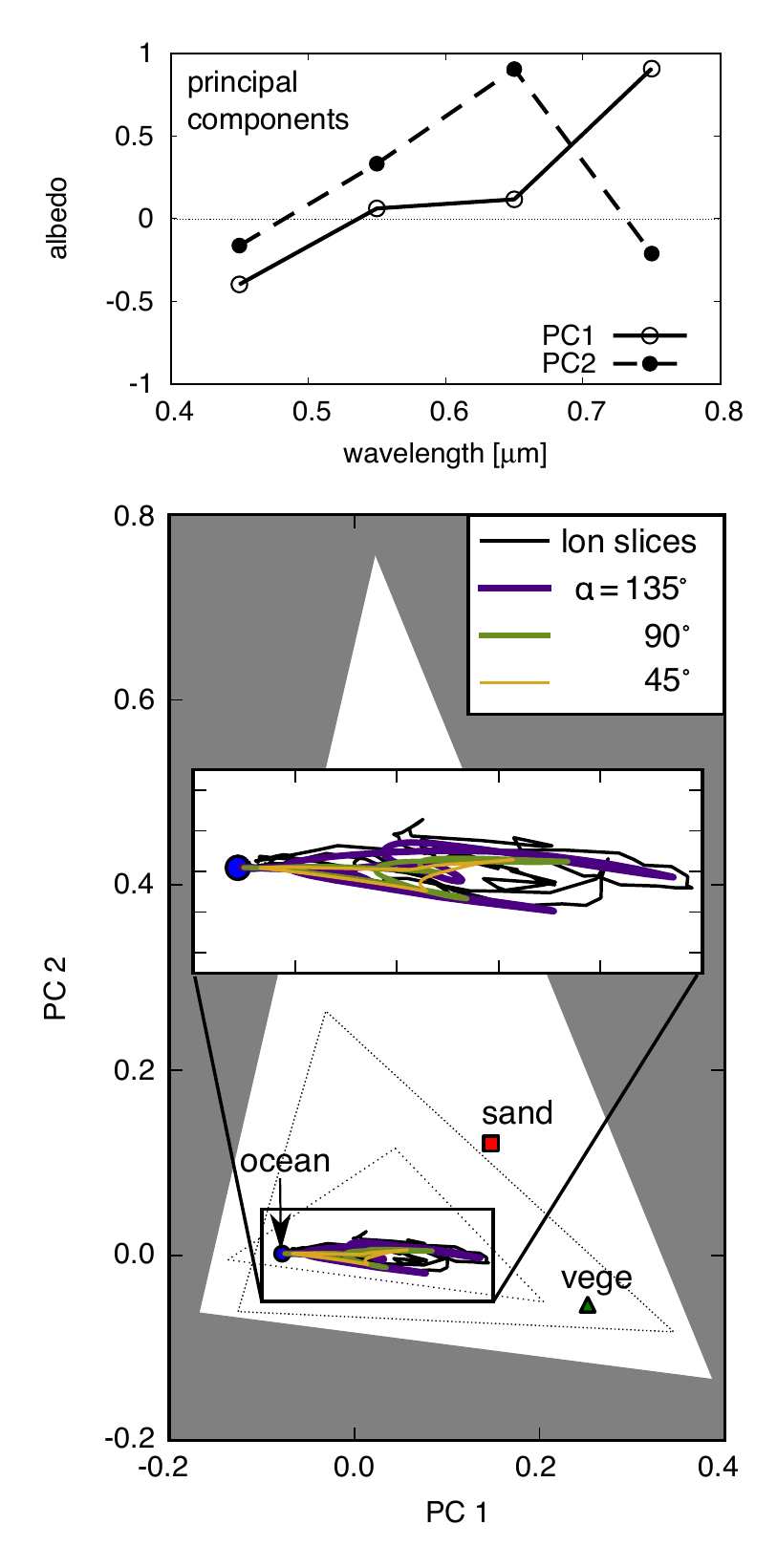}
    \end{center}
    \caption{\emph{Upper panel:} principal components of the light curves at $\alpha = 90^{\circ }$ shown in olive lines in Figure \ref{fig:mockdata}. \emph{Lower panel:} trajectories of the four-band light curves on the PC plane, defined by two PCs shown in the upper panel, in the cases of $\alpha = 135^{\circ }$ (thick indigo line), $\alpha = 90^{\circ }$ (olive line), and $\alpha = 45^{\circ }$ (thin gold line). The points indicate the input albedo spectra of an ocean (blue circle), sand (red square), and vegetation (green triangle) on the PC plane. Points in the gray region violate the condition (\ref{eq:tilde_s_range}); specifically, the left, right, and bottom boundaries are set by $s_{k,4} > 0$, $s_{k,1} > 0$, and $s_{k,3}> 0$, respectively. 
The dotted lines are random triangles that could be solutions (see the text). }
    \label{fig:trajectory}
\end{figure}

From the mock light curves prepared in Section \ref{s:mockdata}, 
we extract two PCs through PCA, with other components having virtually zero contributions, consistent with the three input surface types.  
The PCs extracted from the light curves at $\alpha = 90^{\circ }$ (olive lines in Figure \ref{fig:mockdata}) are presented in the upper panel of Figure \ref{fig:trajectory}. 
For later use, we denote these PCs by $V_{nj}$, where $n=1$ or $2$,  corresponding to the first and second PC. 
The PC spaces (plane) of the light curves at different phases are the same,  since we use the same three surface spectra as inputs, while the individual PCs may be rotated. 

Using the PCs, $V_{nj}$, and the time average of the colors, $\bar d_j$, in the case of $\alpha = 90^{\circ }$, the light curves are projected onto the PC plane by:
\begin{equation}
d_{ij} = \sum_n U_{in} V_{nj} + \bar d_j
\end{equation}
where $U_{in}$ represents the time-dependent trajectory on the plane. 
The trajectories of the light curves at different orbital phases are shown by the colored lines in the lower panel of Figure \ref{fig:trajectory}. 
The color excursions are greater at a larger phase angle (crescent phase), because less surface area is averaged together. 

Likewise, the albedo spectra of representative surface types may be projected onto the PC plane by
\begin{equation}
s_{kj} = \sum_n t_{kn} V_{nj} + \bar d_j ,
\end{equation}
where $t_{kn}$ are the coordinates of the albedo spectra on the plane.
The circle, square, and triangle in the figure indicate the input albedo spectra of ocean, sand, and vegetation, respectively, projected onto this plane. 
As described above, the trajectory of the observations is always in the convex hull (triangle in this case) of the surface albedo spectra. 

\subsection{Degenerate Solutions}
\label{ss:degeneracy}

When it comes to the inverse problem of estimating surface spectra given the trajectory/-ies, {\it any} set of $\{ {\bm s}_k \}$ that encloses the data points $\{{\bm d}_i\}$ in the PC plane can be a solution of Equation (\ref{eq:tilde_d_f_ast_s}) subject to the conditions (\ref{eq:tilde_f_range}) and (\ref{eq:tilde_f_sum}). 
Two such examples are shown by the dotted lines in Figure \ref{fig:trajectory}. 
Note that the associated matrix, $\ftilde _{ik}$, can always be found. 

Additional constraints come from the condition (\ref{eq:tilde_s_range}) that the albedo of each surface be between 0 and 1 in each band. 
In Figure \ref{fig:trajectory} any points in the shadowed region are rejected based on this condition: specifically, the left, right, and bottom boundaries are set by $s_{k,4}> 0$, $s_{k,1}> 0$, and $s_{k,3}> 0$. 
In other words, surfaces lying in the forbidden gray regions have negative albedos in the fourth, first, and third photometric bands. 
While in this particular example the permitted region happens to be a triangle, the shape can differ depending on the location of the PC plane relative to the albedo boundaries. 
Since each of the four bands has lower and upper bounds on albedo, the physically allowed region of color space is a 4-dimensional hypercube, also known as a tesseract. Depending on the orientation of the PC plane with this tesseract, the allowed region may have a variety of geometries. 
In general, the allowed region is an $N$-dimensional slice through a $J$-dimensional hypercube. 

The randomly drawn two triangles in Figure \ref{fig:trajectory} also satisfy the condition (\ref{eq:tilde_s_range}) for an albedo, in addition to the conditions (\ref{eq:tilde_f_range}) and (\ref{eq:tilde_f_sum}), as do many others. 
Thus, predicting $\ftilde _{ik}$ and ${\bm s}_k$ from ${\bm d}_i$ is clearly degenerate. 
For example, with a large triangle one can have spectrally interesting surfaces with boring geography (small longitudinal variation in area fractions), while with a small triangle one can have boring surface spectra (different surfaces have nearly the same color) and interesting geography.

A formally equivalent degeneracy is found in Equation (\ref{eq:d_f_s}) coupled with the conditions (\ref{eq:f_range})-(\ref{eq:s_range}). 
Essentially, the term $\sum _k f_{lk} s_{kj}$ represents the average albedo spectra of $l$-th slice. 
Suppose that ideally we can retrieve the average spectra of longitudinal slices from the light curves through $W_{il}$; the black trajectory in the bottom panel of Figure \ref{fig:trajectory} represents the color variation as a function of longitude based on the upper panel of Figure \ref{fig:mockdata}.  
We again have different sets of end-member colors, $\{ {\bm s}_k \}$, that enclose the averaged albedo spectra of longitudinal slices {\it and} are located in the region of physical albedo presented by condition (\ref{eq:s_range}). 
Nevertheless, the excursions of the longitudinal colors are more dramatic than the disk-integrated colors of the light curves, thus the possible solutions are somewhat restricted.

\subsection{``Best Guess'' of the Surface Types}
\label{ss:guess}

\begin{figure*}[tbh!]
   \begin{minipage}{0.33\hsize}
    \begin{center}
\includegraphics[width=\hsize]{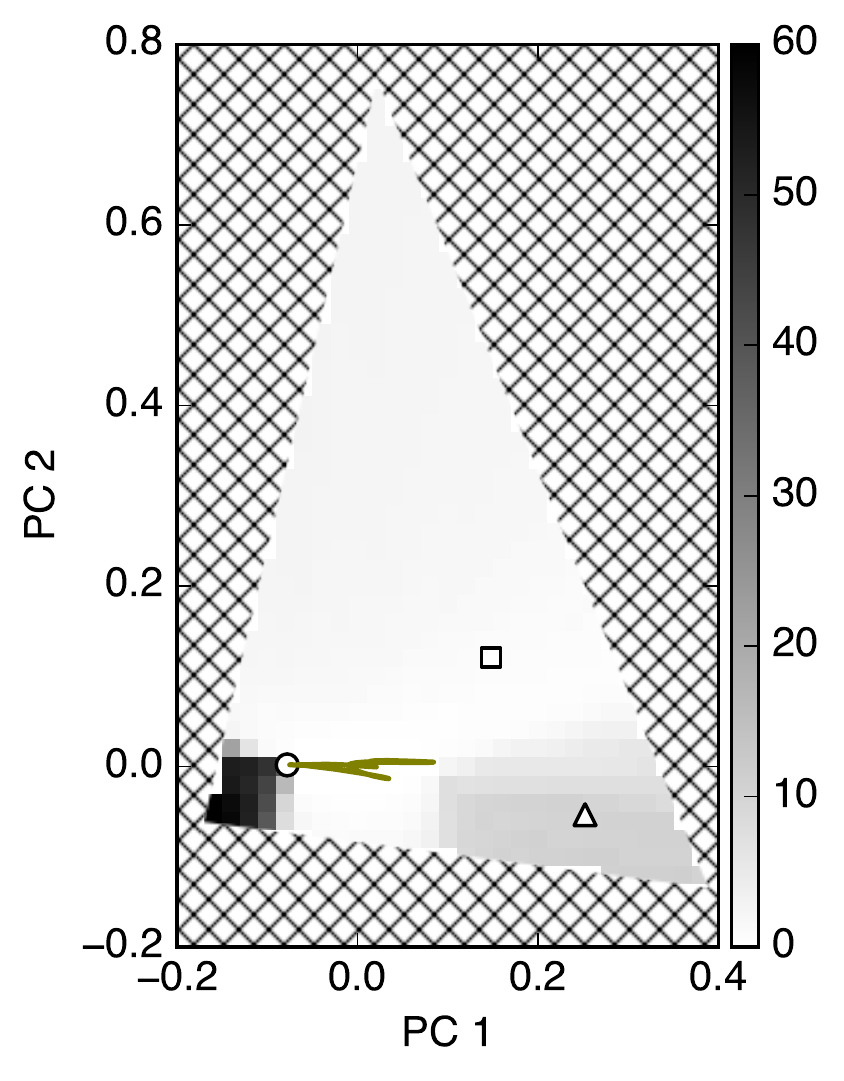}
    \end{center}
     \end{minipage}   
    \begin{minipage}{0.33\hsize}
    \begin{center}
\includegraphics[width=\hsize]{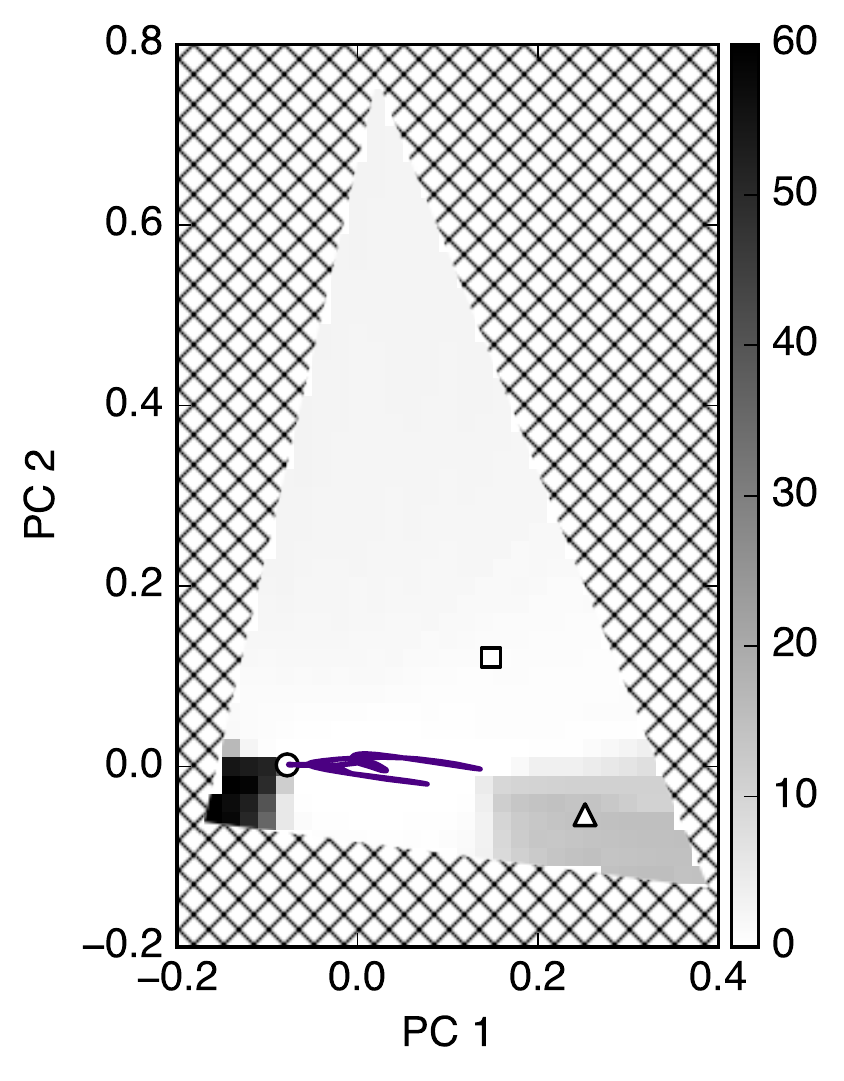}
    \end{center}
     \end{minipage}
   \begin{minipage}{0.33\hsize}
    \begin{center}
\includegraphics[width=\hsize]{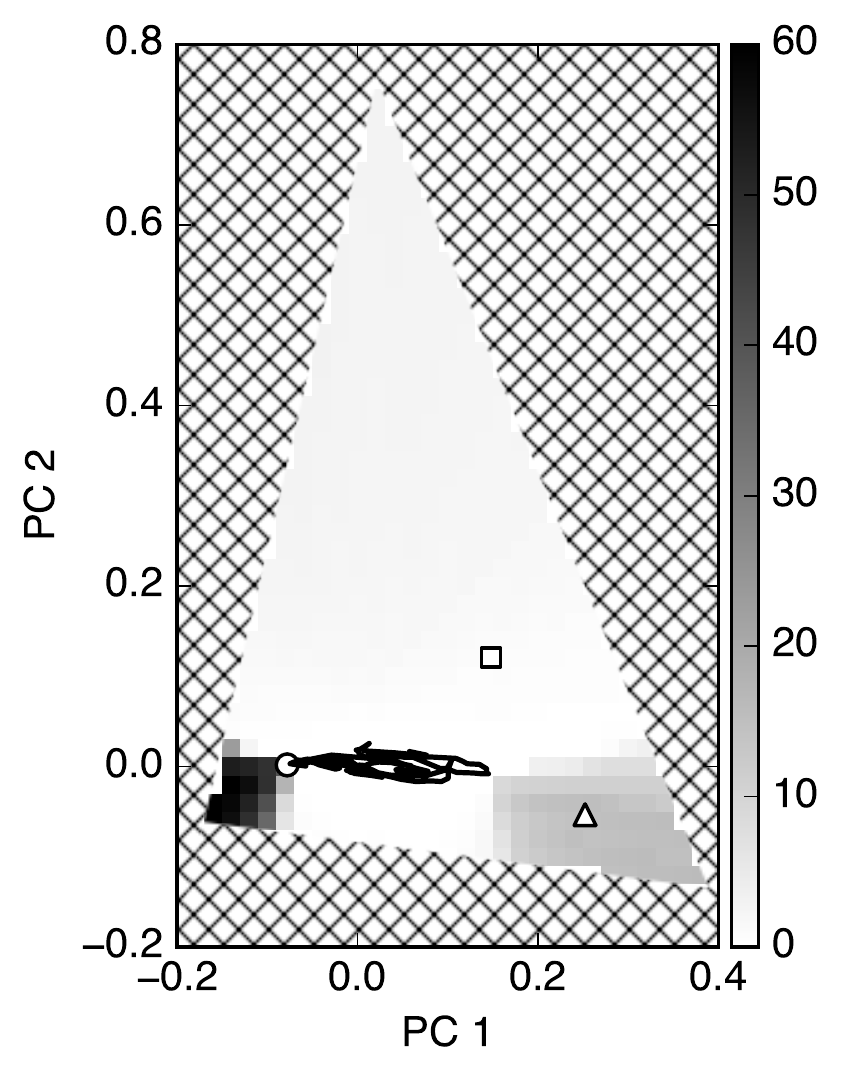}
    \end{center}
     \end{minipage}
    \caption{Probability of surface albedo spectra estimated from the light curves at orbital phases of $\alpha = 90^{\circ }$ (left), $\alpha = 135^{\circ }$ (middle) as well as from the colors of longitudinal slices. The hatched region is not permitted because of negative albedo. Points indicate the input albedo spectra of ocean (circle), sand (square), and vegetation (triangle) on the PC plane. }
\label{fig:PCplane}
\end{figure*}

Given that any three vertices in the PC plane that enclose the trajectory/-ies can be a solution, the choice of the solution depends on the prior probability distribution.  
With no assumptions or information on the geography or spectral albedo of the surface types, it is reasonable to assume that any points in the PC plane except for the forbidden regions are equally likely to correspond to a surface type. 
Despite this uninformative prior, the location of the light curve trajectory allows us to constrain the location of the surface spectra in a probabilistic sense. 
For example, in Figure \ref{fig:trajectory}, in order to enclose the trajectory with three surface types within the permitted region, we must have at least one point near the bottom left corner of the permitted region; this is consistent with the fact that one of the input albedo spectra (ocean) resides there. 

In order to evaluate this more quantitatively, we assume that any combination of three points that enclose the data is equally likely to be a solution for surface end-member colors. We then compute the marginalized probability for the location of one surface type in the PC plane. 
Specifically, we make a grid on the PC plane with an interval of 0.2 and only consider grid points in the permitted region. For each point, we compute the number of combinations of two other points with which it can make a triangle that encloses the trajectory. The number of such triangles is proportional to the probability of there being a surface at that location in color space.  This number is normalized so that the integral over the permitted region is unity. 

The resultant probability distribution is shown in Figure \ref{fig:PCplane}. 
As expected, we find a definite peak near the ocean color. 
The vegetation in the bottom right corner is moderately constrained. 
On the other hand, the very light gray color in the upper part of the triangle indicates that the location of the end-member there (sand) is essentially unconstrained. Note that when one integrates the probability over the upper light gray region, it sums up to about 1/3, implying that one of the end-members should exist ``somewhere'' in the light-gray region.

\subsection{Conditions for Successful Guess\\of Surface Spectra}
\label{ss:guess}

We reiterate that in this framework the success of the surface retrieval critically depends on the relative configuration of the trajectory within the allowed region. 
More specifically, the good constraints on the colors of an ocean appear to be due to the combination of the large covering fraction of the ocean and the location of the ocean color close to a corner in the PC plane (due to low albedo at longer wavelengths). 

\begin{figure}[htb!]
    \begin{center}
    \includegraphics[width=\hsize]{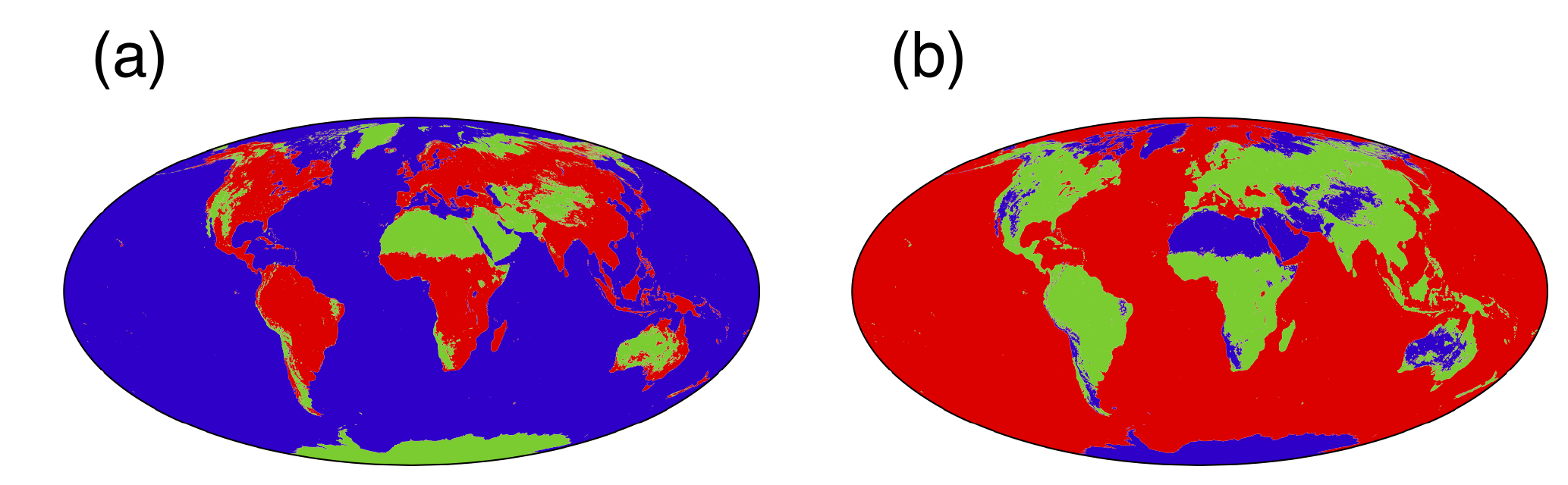}
    \includegraphics[width=\hsize]{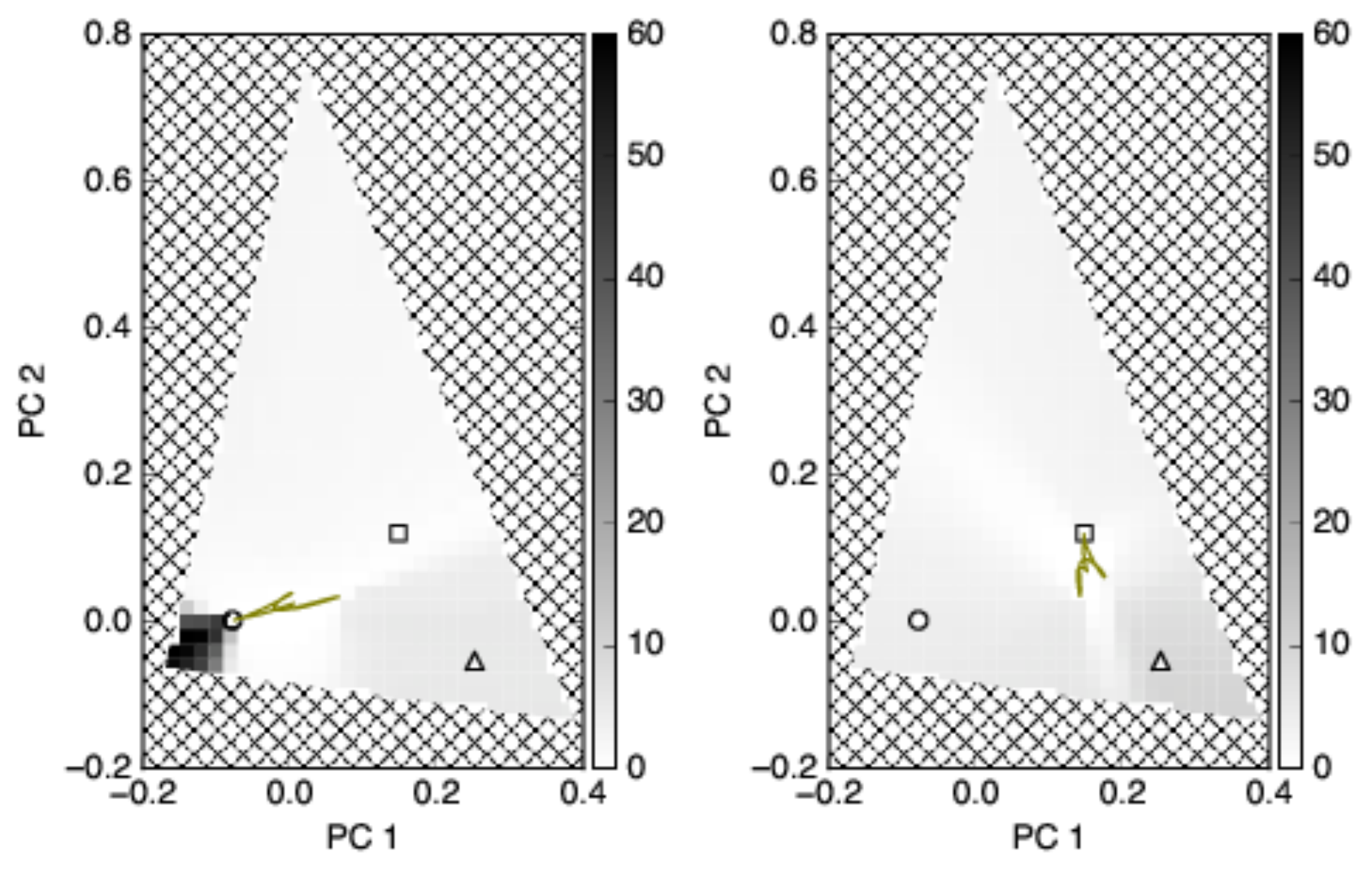}
    \end{center}
    \caption{\emph{Left:} trajectory of the light curves with the geographical maps of  sand and vegetation swapped, and the resultant color contour of the probability for albedo spectra of surface types. \emph{Right:} same as the left panel but with the geographical maps of sand and ocean swapped.}
\label{fig:swap}
\end{figure}

In order to illustrate these two effects and see under what conditions we can faithfully retrieve surface spectra, we create two additional mock light curves, (a) with the geographical maps of sand and vegetation swapped, and (b) with the geographical maps of sand and ocean swapped. 
The left and right panels of Figure \ref{fig:swap} correspond to (a) and (b), respectively. 

When we swap sand and vegetation (left panel), the covering fraction of vegetation becomes smaller, and the trajectory of the light curves does not approach the vegetation end-member. As a result, the constraints on the color of vegetation become weaker. Thus, a large covering fraction of the surface type tends to yield a better constraint on the albedo spectra of that type; this agrees with our intuition.  

Meanwhile, in (b) we swap sand and ocean.  
Although sand is now the dominant component, 
our analysis prefers other locations toward the top of the allowed region, because there is more space there. 
Therefore, a large covering fraction is not a sufficient condition for obtaining strong constraints on the color of a surface. 
Better constraints on a surface spectrum can be obtained if its location on the PC plane is close to more than one boundary of the allowed region. 
In other words, a surface spectrum can be well retrieved if it has a  favorable geography (ideally an entire hemisphere solely covered with that surface) {\it and} it has a favorable albedo spectrum (albedo close to 0 or 1 at as many wavelengths as possible).

\section{Application to EPOXI data}
\label{s:EPOXI}

\begin{figure}[t!]
    \begin{center}
	\includegraphics[width=0.9\hsize]{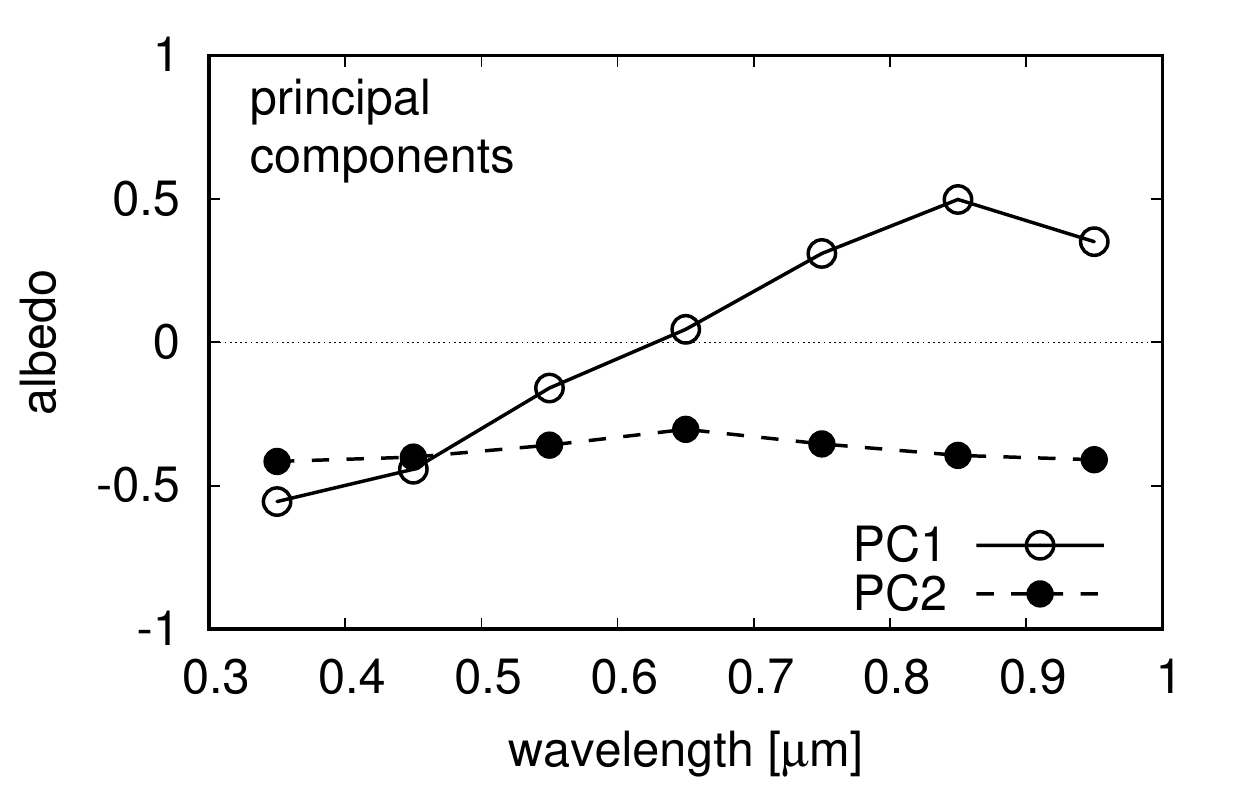}
	\includegraphics[width=0.9\hsize]{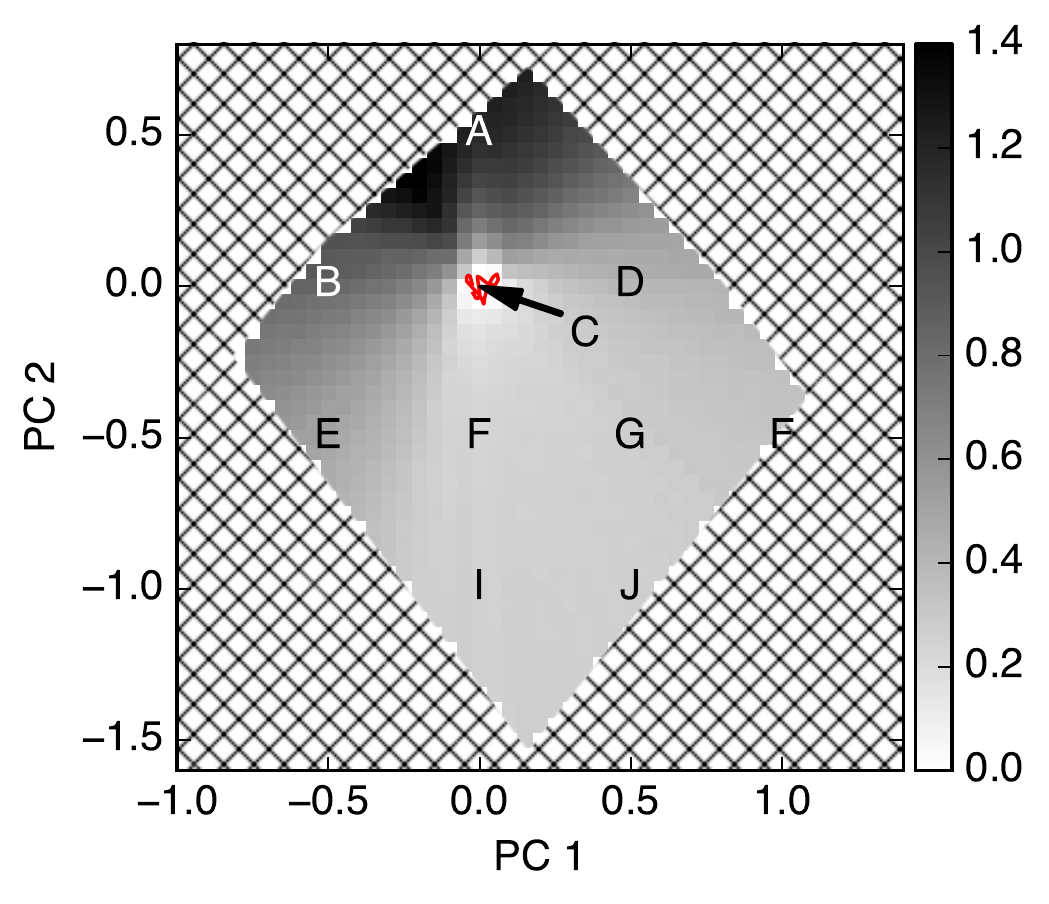}
	\includegraphics[width=0.89\hsize]{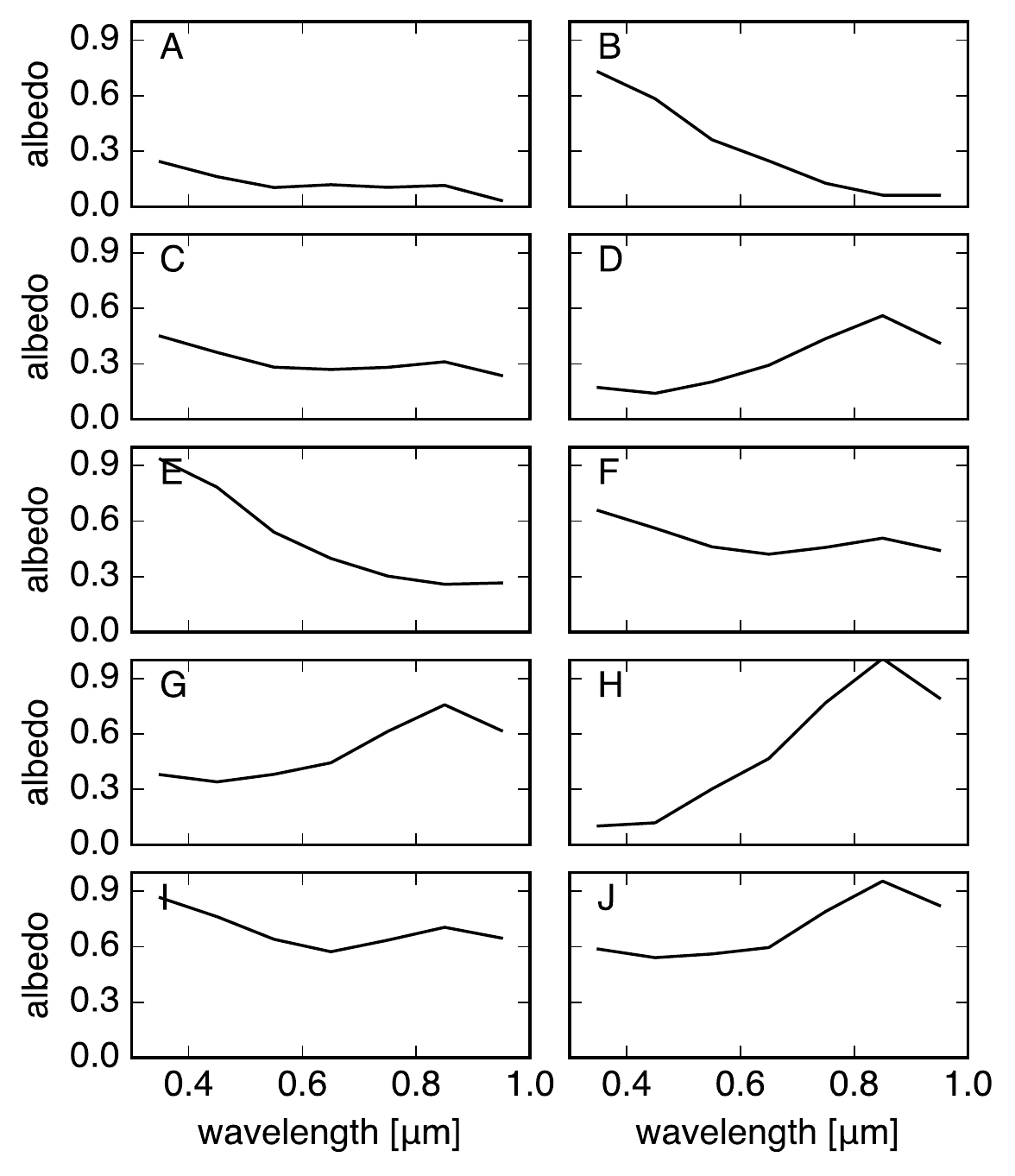}
    \end{center}
    \caption{\emph{Top panel:} the two dominant principal components from EPOXI June data \citep{Livengood2011}. \emph{Middle panel:} color contour similar to Figure \ref{fig:PCplane}, but based on EPOXI June data. The red line shows the light curve trajectory. The hatched region is not permitted because of the physicality constraints for albedos. \emph{Bottom panels:} spectra corresponding to the locations (A-J) shown in the upper panel. The ocean, sand, and clouds appear to be close to A, D, and F, respectively. }
\label{fig:EPOXI}
\end{figure}

In this section, we apply our procedure to the multi-band light curves of Earth obtained by the EPOXI mission. 
Specifically, we use the same datasets as \citet{Cowan2013}, i.e., the seven-band diurnal light curves observed in June 2008 \citep{Livengood2011}.  
We obtain the two dominant PCs shown in the top panel of Figure~\ref{fig:EPOXI}, which suggests at least three spectrally distinct surface types, consistent with \citet{Cowan2013}. 
Using these two PCs, we perform the same analysis described in the previous section. 

The middle panel of Figure \ref{fig:EPOXI} shows the trajectory of the light curve on the PC plane, with the hatched region being forbidden by the physicality constraints for albedo. 
While the permitted region appears to be close to a square, it is in fact a heptagon, bounded by seven inequalities (2-dimensional slice through a 7-dimensional hypercube). 
Overall, the upper boundaries come from the requirement that albedo cannot be negative at any wavelength, i.e., $s_{kj}>0$, while the lower boundaries comes from the requirement that albedo should be less than unity, i.e., $s_{kj}<1$. 

In principle, any three or more points that enclose the trajectory can be solutions for the surface albedo spectra. The example albedo spectra corresponding to different locations in the permitted region are shown in the lower panels of Figure \ref{fig:EPOXI}. This library conservatively represents the range of possible surface spectra. 

Going one step further, we evaluate the preferred surface spectra following the procedure discussed in Section \ref{ss:guess}. We assume that there are three spectrally distinct surface types. The grayscale in the middle panel of Figure \ref{fig:EPOXI} indicates the probability distribution equivalent to those in Figures \ref{fig:PCplane} and \ref{fig:swap}, which is simply related to the number of possible combinations each location can make to enclose all of the data points. 
In this case, the trajectory of planetary color is far from the boundary because the color variations of the real Earth are muted by the cloud cover. 
Consequently, we do not see three peaks. 
Nevertheless, we are likely to find one surface type in the upper part of the allowed region, near label A, which can be interpreted as ocean.

\section{Discussion}
\label{s:discussion}

\subsection{Confounding Factors}
\label{ss:confonting_factors}

\subsubsection{Uncertain Planetary Radius}
\label{sss:uncertain_radius}

So far, we have assumed that the data can be expressed in terms of apparent albedo. 
In reality, the primary observables from direct imaging observations are the intensities of the planet and the star. 
In order to convert them to apparent albedo, we need to know the orbital distance from the star, the phase angle, and the planetary radius. 
While the former two may be constrained from multi-epoch observations, the radius continues to be severely degenerate with albedo. 
How would the unknown planetary radius affect our retrieval? 

Not knowing the planetary radius and hence the normalization of the planetary albedo, we cannot put a tight upper limit on the surface albedo spectra, $s_{kj}$. 
Thus, the condition (\ref{eq:tilde_s_range}) or (\ref{eq:s_range}) is essentially  reduced to $0 \leq s_{kj}$. 
How much this limits our estimation depends on the configuration of the PC space in the $J$-dimensional color space. 
For example, in the case of our mock data, the three significant constraints actually come from $0 \leq s_{kj}$ (see Figure \ref{fig:trajectory} and its caption), therefore the unknown normalization of planetary albedo does not affect the permitted region. 
Thus, the constraints of the spectral shapes of the surfaces can be obtained (but not the absolute scale). 
On the other hand, in the case of EPOXI data, roughly half of the boundaries of the permitted region originate from the upper limit for albedo, $s_{kj} \leq 1$. 
In this case, not knowing the normalization would change the permitted region itself. 
Nevertheless, because the upper boundaries still come from $0 \leq s_{kj}$, the inference that at least one surface type is present near the upper left corner will hold. 

\subsubsection{Additional Coplanar Surface Spectra}

The retrieval would become more complicated if the actual number of surface types were larger than the dimension of the PC space plus 1 ($K > N + 1 $). 
For example, if there were four important surface types in EPOXI rather than three, all in the PC plane we obtained, then we would have to consider a tetragon to enclose all of the data rather than a triangle, thus affecting our estimates of surface spectra. 
The greater $J$ is, the more probable it is for important surface colors in the $J$-dimensional space to be affinely independent (i.e., to make a simplex).
Thus, obtaining light curves in as many bands as possible will help disentangle the surface colors and reduce the uncertainty in the number of surface types.

\subsubsection{Deviation from Lambert's Law}
\label{ss:deviate_Lambert}

We have assumed that the surface scattering obeys Lambert's law, but in reality they are not perfectly Lambertian. 
In particular, scattering by surface liquid water is characterized by specular reflection and the reflectivity increases when the incident light is grazing, or equivalently at crescent phase \citep[e.g.,][]{Williams2008,Robinson2010,Robinson2014}. 
Scattering by cloud/haze layers exhibits prominent forward scattering, which also becomes evident at crescent phase \citep[e.g.,][]{Robinson2010}. 
For surfaces with strongly anisotropic scattering, apparent albedo can even exceed unity. 
Furthermore, the albedo spectrum of a surface overlaid by an atmosphere changes with phase angle.  
Considering these effects, the trajectories of the light curves at different phases do not have to reside in the same PC space. 
These light curves can be analyzed independently, and the components varying with phase would give us insights into the anisotropic scatterers.

\subsection{Enhancing Constraints}
\label{ss:enhancing_constraints}

\subsubsection{Power of Many Bands}

As we discussed in Section \ref{ss:guess}, constraints on surface spectra rely on the relative configuration between the light curve trajectory and the boundaries of the physically permitted region, namely that the albedo of any surface must be between 0 and 1 at all wavelengths. 
It depends on geography and the albedo spectra of the surface types, neither of which we can directly change. 
However, we can potentially increase the number of boundaries of the permitted region by increasing the number of photometric bands, and some of them may have tighter constraints than others. 
Obtaining light curves in as many bands as possible will therefore be beneficial. 

In particular, observing at wavelengths where one or more surface types have albedos close to 0 or 1 constrains the possible albedos more tightly. 
Albedos close to 0 (i.e., the condition of $0 \le s_{kj}$) are more useful because they do not depend on the uncertain planetary radius. 
In reality, the albedo of ocean is close to 0 at longer wavelengths toward the near infrared, that of sand is small at 0.4 $\mu $m and shortward, that of vegetation is close to 0 at 0.7 $\mu$m and shortward, and that of H$_2$O snow is close to 1 in the visible and closer to 0 at the 1.5 $\mu$m and 2 $\mu$m bands. 
Observations at different bands covering these characteristic wavelengths would be useful. 

\begin{figure*}[hbt!]
   \begin{minipage}{0.33\hsize}
    \begin{center}
\includegraphics[width=\hsize]{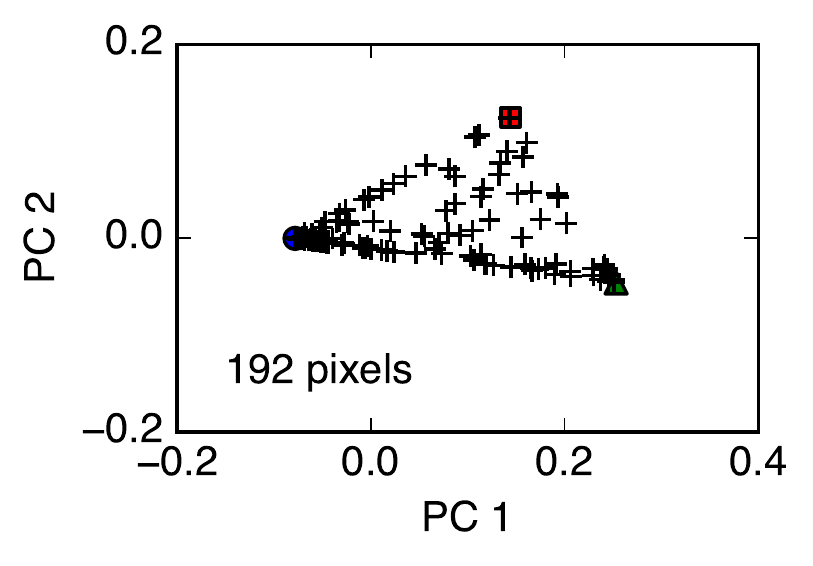}
    \end{center}
     \end{minipage}   
    \begin{minipage}{0.33\hsize}
    \begin{center}
\includegraphics[width=\hsize]{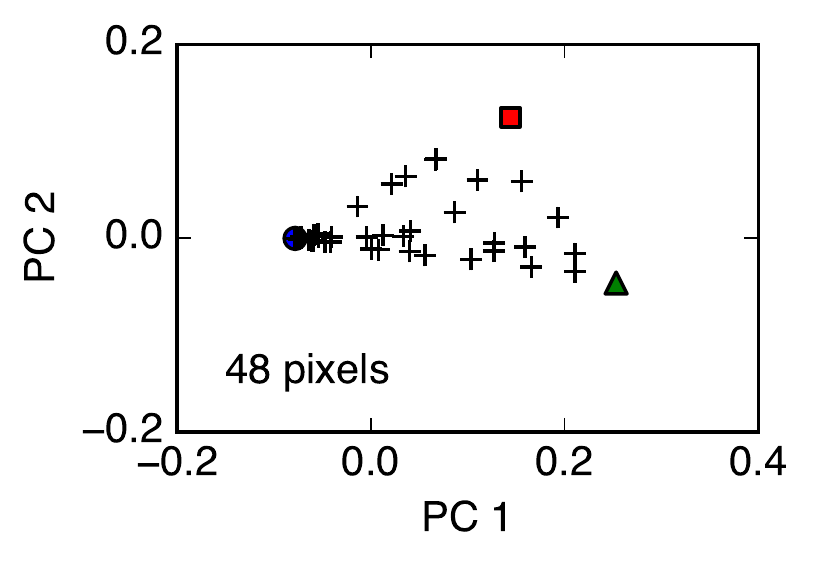}
    \end{center}
     \end{minipage}
   \begin{minipage}{0.33\hsize}
    \begin{center}
\includegraphics[width=\hsize]{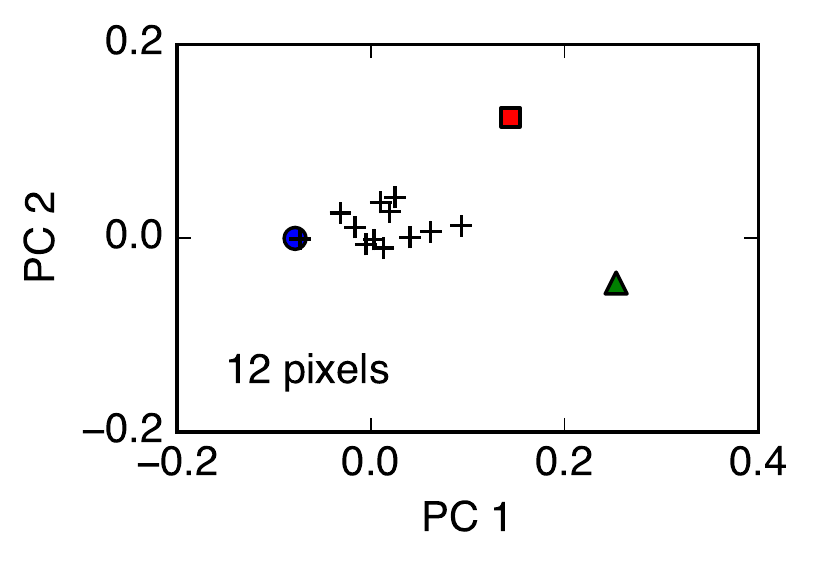}
    \end{center}
     \end{minipage}
    \caption{Making the map coarser from the upper panel of Figure \ref{fig:mockdata}. Pixelating the surface into 192 (left), 48 (middle), and 12 (right) equal-area pixels using HealPix, we plot the synthesized pixel colors with crosses on the same PC plane as Figures \ref{fig:trajectory} to \ref{fig:swap}.  The triangular shape of the locus of points becomes apparent with as few as 48 pixels. Thus, resolving the surface colors in 2 dimensions eases the challenge to identify surface types. }
\label{fig:lowresolution}
\end{figure*}

\subsubsection{Utilizing Light Curves at Different Phases}

Light curves at different orbital phases are useful for spectral unmixing  because they are sensitive to different locations on the planet, exploring the PC space in different directions. 
As a result, the volume of color space where the solution can exist will be narrowed.  
In particular, the diurnal light curves at thinner phases tend to exhibit larger color variations, giving tighter constraints (but see the caveat in Section \ref{ss:deviate_Lambert}). 

In addition, diurnal light curves at different orbital phases in principle allow us to map the color in two dimensions \citep{Kawahara2010,Kawahara2011,Fujii2012}. 
We could then use the spectra of various surface patches resolved in two  dimensions to identify different surface types, in the same spirit as that of estimating the end-member colors from the colors of the longitudinal slices. 
In fact, this ``map first'' approach is likely more appropriate with 2-dimensional maps than with longitudinal maps. 
In the diurnal light curves at a single phase, colors of very distant pixels beyond the correlation length of the geography are mixed together, even at crescent phases---for example, the colors of the Arctic, tropics, and Antarctic are mixed in an equatorial observation, no matter how extreme the orbital phase is. 
As a result, the colors of longitudinal slices never approach the colors of vegetation or sand regardless the number of slices, as shown in the right panel of Figure \ref{fig:PCplane}. 
The situation is different when we gradually lower the resolution of the 2-dimensional maps. 
Figure \ref{fig:lowresolution} shows the colors of HEALPix pixels \citep{Gorski2005} with varying resolutions: from left to right, we change the resolution from 192 pixels, to 48 pixels, to 12 pixels, starting from the map in Figure \ref{fig:mockdata}. 
The triangular shape of the locus of points becomes apparent with as few as 48 pixels. 
Thus, resolving the surface colors in two dimensions will potentially ease the challenge of identifying surface types.

\subsubsection{Invoking Additional Assumptions}

Occasionally, we may want to adopt additional constraints on the behavior of $\ftilde _{ik}$ ($f_{lk}$) or $s_{kj}$ based on prior knowledge. 

For example, we may assume that the covering fractions of surface types vary  ``smoothly'' as a function of time or longitude. 
Such an assumption could be implemented via a Gaussian process \citep[e.g.,][]{Rasmussen2005}, i.e., the area fractions at different times/longitudes have a joint gaussian distribution with the relevant correlation length. 
We must exercise caution because the actual covering fractions of surface types are not necessarily Gaussian processes and can exhibit sharp boundaries. 

We could also consider a regularization on albedo spectra as a function of wavelength. 
This can be regarded as a prior probability distribution on the PC plane. 
For example, in Figure \ref{fig:PCplane}, albedo spectra corresponding to the upper part of the permitted region exhibit a dramatic decrease in albedo between 0.65~$\mu $m and 0.75~$\mu $m, which may seem implausible. 
However, the albedo spectrum of vegetation in fact has a sharp feature, the red edge; similar, unfamiliar features may exist on other planets. 
In addition, with an atmosphere, absorption by atmospheric molecules can also produce sharp changes in albedo spectra.

\section{Conclusion}
\label{s:conclusion}

In this paper, we revisited the problem of estimating the albedo spectra of major surface types and their distributions from disk-integrated colors of exoplanets. 
We pointed out the inherit degeneracy that makes it impossible to find unique solutions. 
Despite the degeneracy, the following physical conditions narrow down the possible solutions for surface albedos, on the assumption of Lambertian surfaces:
the actual surface albedo spectra in the $J$-dimensional color space are 
(1) to be located in the PC subspace, 
(2) to enclose all of the data points of the disk-integrated colors, and (3) to be in the $J$-dimensional hypercube, bounded by 0 and 1 for all axes. %
We demonstrated using both a simplified toy model of Earth and the observed data by EPOXI that such constraints point us to the approximate spectrum of ocean. 

In general, we find that the success of the estimates critically depends on the degree of excursions of the time-dependent trajectory of the disk-integrated colors, and its orientation in the color space. %
In other words, the estimates will be better if the covering fraction of each surface type becomes close to 1, as intuition suggests. 
In addition, observations at wavelengths where surface albedos are close to 0 or 1 could improve the estimates. 
As the wavelength coverage/resolution increases, and as the light curves are obtained at more orbital phases, precision for surface estimates will improve.

\acknowledgements

We thank the anonymous referee whose comments helped us greatly improve the clarity of this manuscript. 
We acknowledge the Exo-Cartography workshops supported by the International Space Science Institute (ISSI). 
We thank D. Foreman-Mackey for a useful discussion of parameter retrieval. 
Y.~F. is thankful to Eric Smith and Haru Negami for discussions on the geometrical aspect of the study. 
Y.~F. acknowledges the generous support from the Universities Space Research Association through an appointment to the NASA Postdoctoral Program at the NASA Goddard Institute for Space Studies. 
Y.~F. is also supported by the NASA Astrobiology Program through the Nexus for Exoplanet System Science. 
J.~L.~Y. is supported by the NASA Astrobiology Institute's Virtual Planetary Laboratory under Cooperative Agreement number NNA13AA93A. 
N.~B.~C. is supported by the McGill Space Institute and l'Institut de recherche sur les exoplan\`etes.  

\bibliography{ref}

\end{document}